# A CW superconducting linac as the proton driver for a medium baseline neutrino beam in China[*]

LI Zhi-Hui（李智慧）[1,2], TANG Jing-Yu （唐靖宇）[2]

1 Institute of Nuclear Science and Technology, Sichuan University, Chengdu, 610049, China
2 Insitute of High Energy Physics, CAS, Beijing, 100049, China

**Abstract**. In a long-term planning for neutrino experiments in China, a medium baseline neutrino beam is proposed which uses a CW superconducting linac of 15 MW in beam power as the proton driver. The linac will be based on the technologies which are under development by the China-ADS project, but with much weaker requirement on reliability. It is composed of a 3.2-MeV normal conducting RFQ and 5 different types of superconducting cavities. The nominal design energy and current are 1.5 GeV and 10 mA, respectively. The general considerations and preliminary results on the physics design will be presented here. In addition, the alternative designs such as 2.0 GeV and 2.5 GeV as they may be required by the general design can be easily extended from the nominal one.

## 1  Introduction

A muon-decay medium baseline neutrino beam named as MOMENT has been proposed to study leptonic CP violation in a decade from now [1-2]. The proton driver is defined as a CW superconducting linac with beam power of 15 MW, and the beam energy is still under optimization among 1.5, 2.0 and 2.5 GeV depending on the efficiency of muon production and the cost. There are a lot of technical challenges in building such a high beam power proton linac and intensive R&D efforts are needed. Fortunately, in 2011, China launched the Accelerator Driver System project (China-ADS or C-ADS) [3], and it is foreseen that most of the critical technical problems and solutions about the 15-MW linac can be solved or proven in the near future in the frame of C-ADS project. Therefore, one of the most important strategies in the proton driver design is to apply the same technology as the C-ADS linac, for example using same types of cavities, but take away the excessive safety margin and tolerant design applied in the C-ADS linac, which are to meet very stringent requirement on reliability and stability for ADS applications [4]. In the following we will present a design for the MOMENT proton driver, which corresponds to the highest energy, namely 2.5 GeV in energy and 10 mA in current, though the nominal design uses 1.5 GeV and 10 mA.

## 2  Design considerations and lattice design for subsections

The front-end is defined as 10 MeV in energy and it is totally the same as the Injector Scheme I [3] of the C-ADS linac as shown in Figure 1. It is composed of an ECR ion source with 35 kV extraction voltage, a low energy beam transport line (LEBT), a 3.2 MeV Radio-Frequency Quadrupole accelerator (RFQ), a medium energy transport beam line (MEBT) and a cryomodule with 12 superconducting spoke cavities with geometry beta 0.12 and 11 superconducting solenoids inside. The fabrication of the normal conducting part of the injector has finished and will start commissioning from early 2014. The prototypes of the superconducting spoke cavity have been tested and the massive production is underway. The compact lattice structure is applied for the superconducting section since as the study shows the longer period length will decrease the longitudinal acceptance and the stability of the beam dynamics [5].

---
[*] Supported by National Natural Science Foundation of China (11375122, 91126003)
1)  E-mail: lizhihui@scu.edu.cn

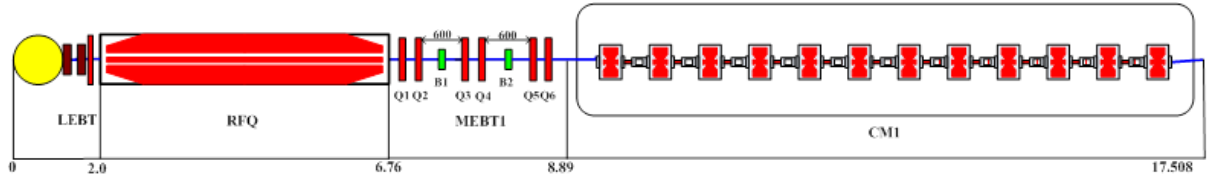

Fig.1 Layout out of the front end of the driver linac

In the main linac part, the same types of cavities as the ones used for the C-ADS main linac, namely spoke021, spoke040, ellip063 and ellip082 are applied and the main properties of the cavities can be found in Ref. [3]. One of the most significant differences between the proton driver and the C-ADS linac is that the full potential of the cavities in the main linac is employed in the design. For the C-ADS linac, only the 3/4 of the cavity capabilities is used in order to realize the local compensation for cavity failures. The general design criteria of high power linacs [6] are followed here, thus the lattice structures have to be redefined in order to satisfy the phase advance law, namely the zero-current phase advance per period should be less than 90 degrees and zero-current phase advance per meter should change smoothly. Both of them can be met by properly setting the period lengths and the transition energies between different sections. The phase advances are shown in Figure 2.

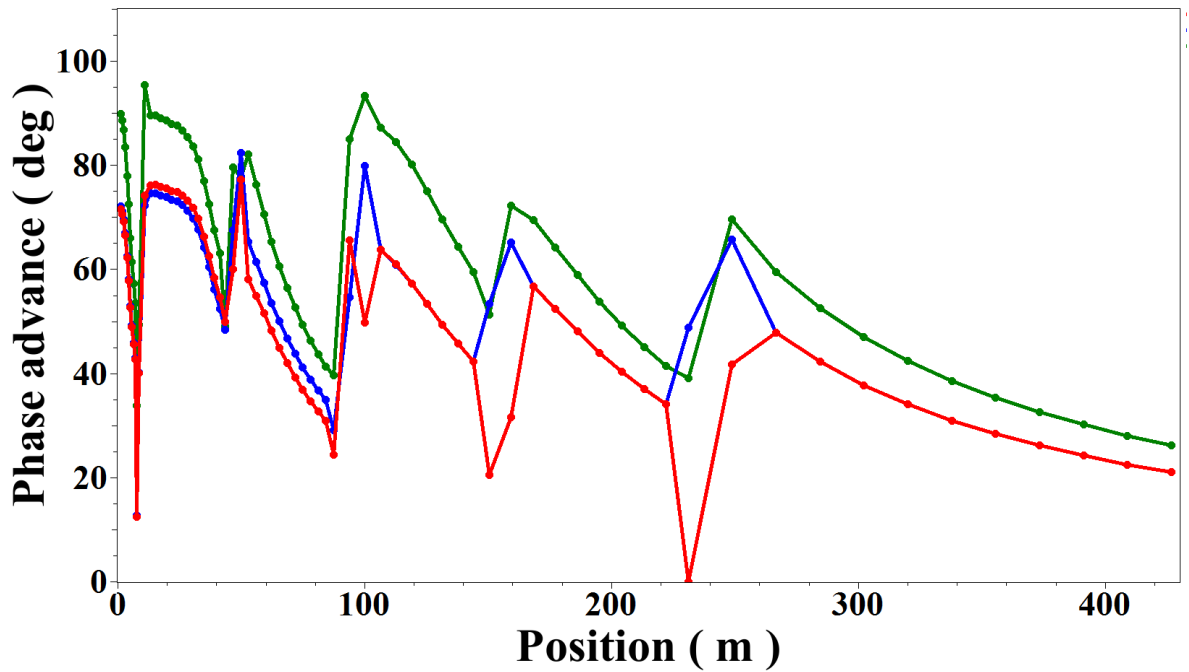

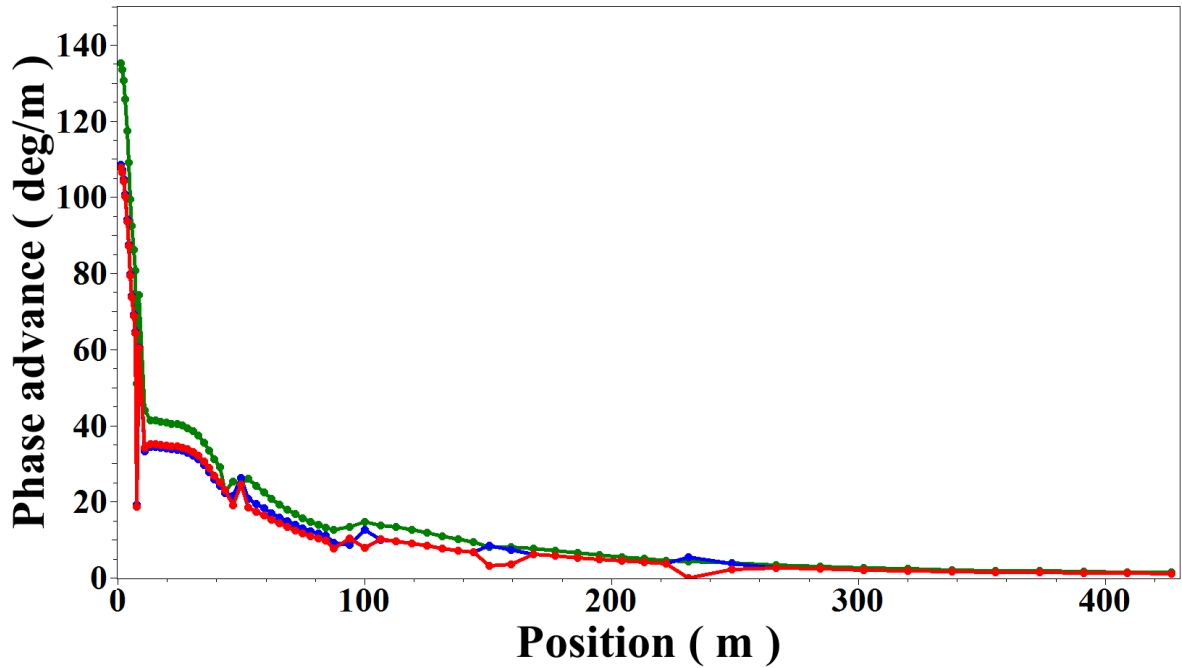

Fig.2 Zero-current phase advances (Upper: phase advance per period, Lower: phase advance per meter, green line: longitudinal direction, red line: x direction, blue line: y direction)

The corresponding lattice structures are shown in Figure 3. There are two and three cavities per period for the Spoke021 section and the Spoke040 section, respectively, and a superconducting solenoid per period is applied for transverse focusing. For the elliptical sections with energy lower than 1 GeV, normal conducting quadrupole doublets are used for transverse focusing. For the Ellip063 section, there is only three cavities in one period labelling with $R^3FD$ (R for Resonator, F for focusing quadrupole, D for defocusing quadrupole) in order to make the phase advance smoothly transferred between the Spoke040 section and the Ellip063 section. For the Ellip082 section, there are five cavities in one period labelling with $R^5FD$. For the Ellip082 section with energy higher than 1 GeV, the $R^5FR^5D$ structure is applied as the transverse focusing structure to double the phase advance per period.

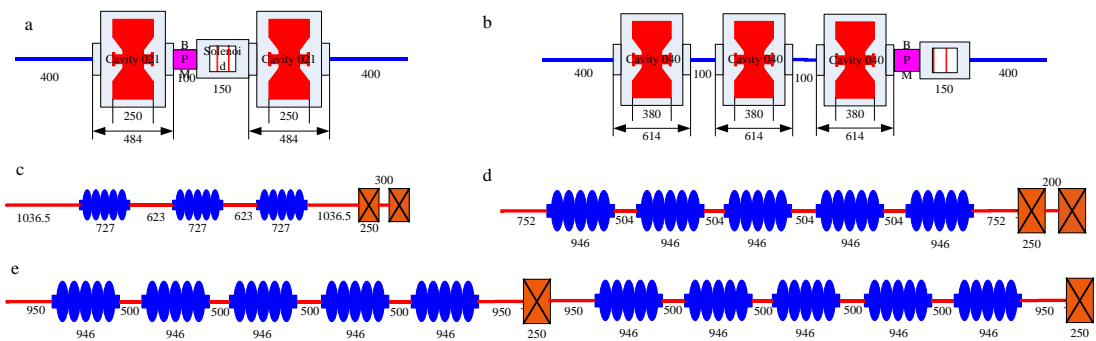

Fig.3 Lattice structure of the main linac(a: Spoke021, b: Spoke040 section, c: Ellip063 section, d: Ellip082-A section (<1 GeV), e: Ellip082-B section (>1 GeV))

Based on the experience on the C-ADS linac design, relatively larger absolute synchronous phases are applied in order to obtain a larger longitudinal acceptance as shown in Figure 4. The ratio between the transverse and longitudinal focusing strengths is determined by making sure that the working points along the linac periods are located in a resonance-free region in the Hofmann chart [7]. Here it is set as 0.8 according to the ratio of the longitudinal and transverse emittances at the exit of the RFQ, so that the equipartition condition can be approximately satisfied. Figure 5 shows the location of the working points in the Hofmann chart.

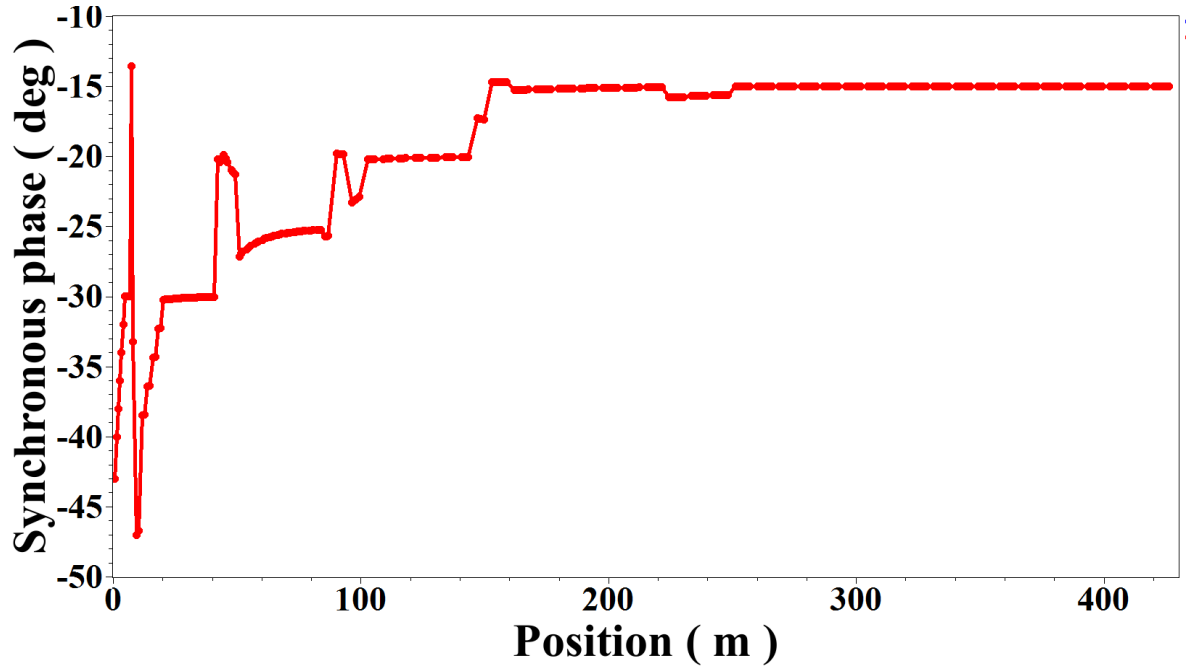

Fig. 4 the synchronous phase variation along the driver linac

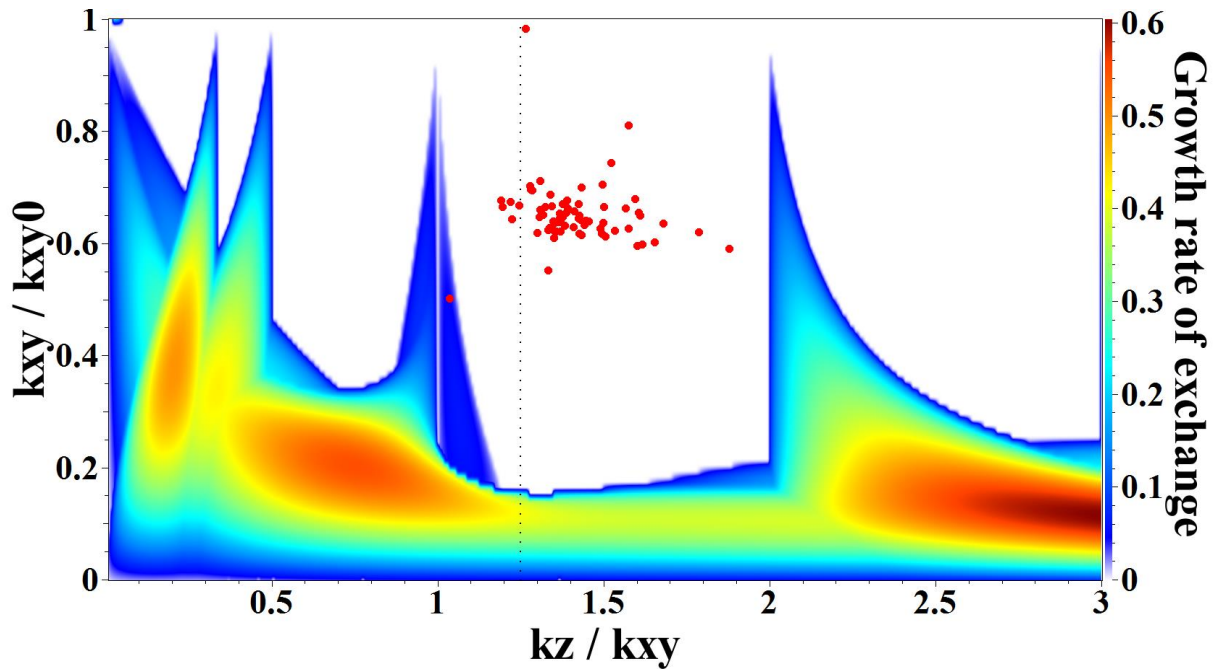

Fig.5 Footprint of the working points in the Hofmann chart

The main parameters of the proton driver are summarized in table 1.

**Table 1.** The main parameters of the proton driver linac

|  | Spoke012 | Spoke21 | Spoke040 | Ellip063 | Ellip082-1 | Ellip082-2 | Ellip082-3 | Total |
|---|---|---|---|---|---|---|---|---|
| Energy (MeV) | 10 | 40 | 160 | 409 | 1000 | 1500 | 2500 | 2500 |
| Cavity number | 12 | 32 | 42 | 30 | 45 | 45 | 70 | 276 |
| Focusing structure | RS | RSR | $SR^3$ | $FDR^3$ | $FDR^5$ | $FR^5DR^5$ | $FR^5DR^5$ |  |
| Total leng. (m) | 8.768 | 43.456 | 87.444 | 150.444 | 230.994 | 293.544 | 418.144 | 418.1 |
| Section leng. (m) | 8.768 | 34.688 | 43.988 | 63.000 | 80.550 | 62.550 | 124.600 |  |
| CM Number | 1 | 8 | 7 | 10 | 9 | 9 | 14 | 58 |
| Syn. Phase | -43--30 | -45--30 | -25 | -20 | -15 | -15 | -15 |  |

## 3 Multi-particle simulations

Multi-particle simulations have been performed to verify the validation of the design. The RFQ is simulated by ParmteqM [8] and the simulated output beam parameters are used as the input parameters and a 4σ Gaussian distribution with 100,000 particles with 10 mA current are generated as the input distribution for the rest of the linac. For all the superconducting cavities, the 3-D field maps based on the cavity electromagnetic designs are used. For the transverse elements, the hard-edge approximation matrices are used and the validation is checked with the field maps. The TraceWin [9] code is used for the simulations after the RFQ.

The simulation results are quite promising. Figure 6 shows the RMS envelopes along the driver linac. We can see that the envelopes change smoothly and the matches between different sections are almost perfect. As the emittance growth may cause halo production and particle loss which are less tolerated in fraction in high-power linacs, it should be strictly controlled. For the proton driver, we can see the emittance growths in the transverse planes are less than 5%, and in the longitudinal one it is less than 10% as shown in Figure 7. The particle distribution in the different phase space projection planes are shown in Figure 8. We can see that the particles are well confined and no phase space distortion happens.

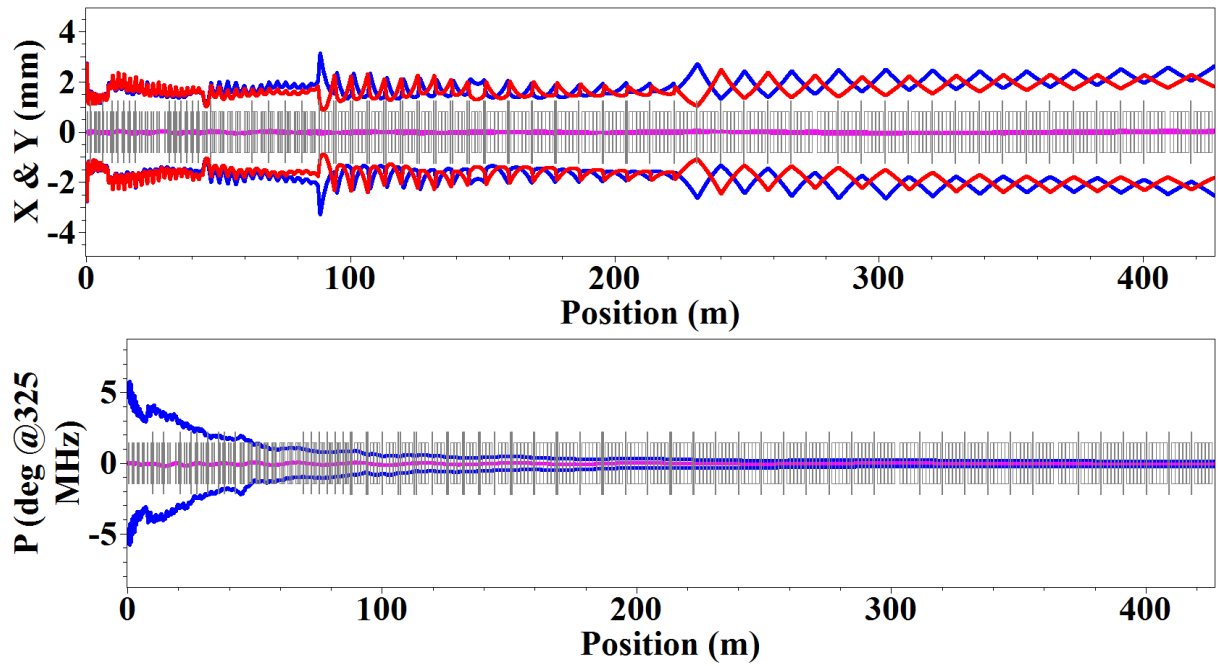

Fig.6 RMS envelop along the linac

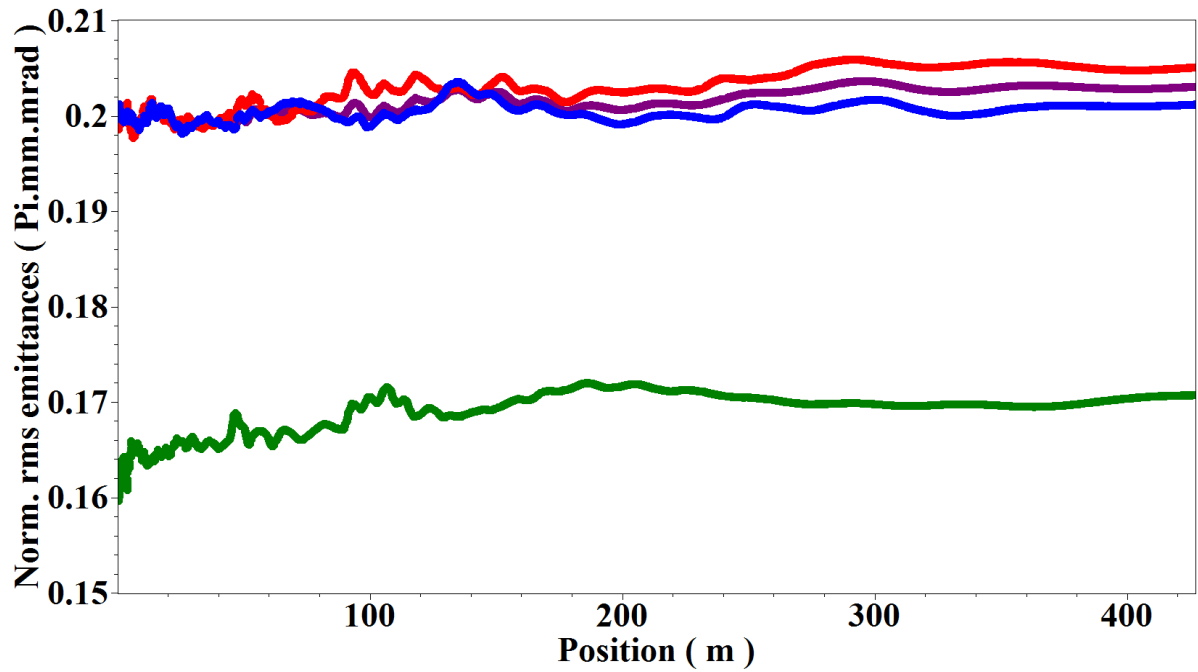

Fig. 7 evolution of the normalized RMS emittances

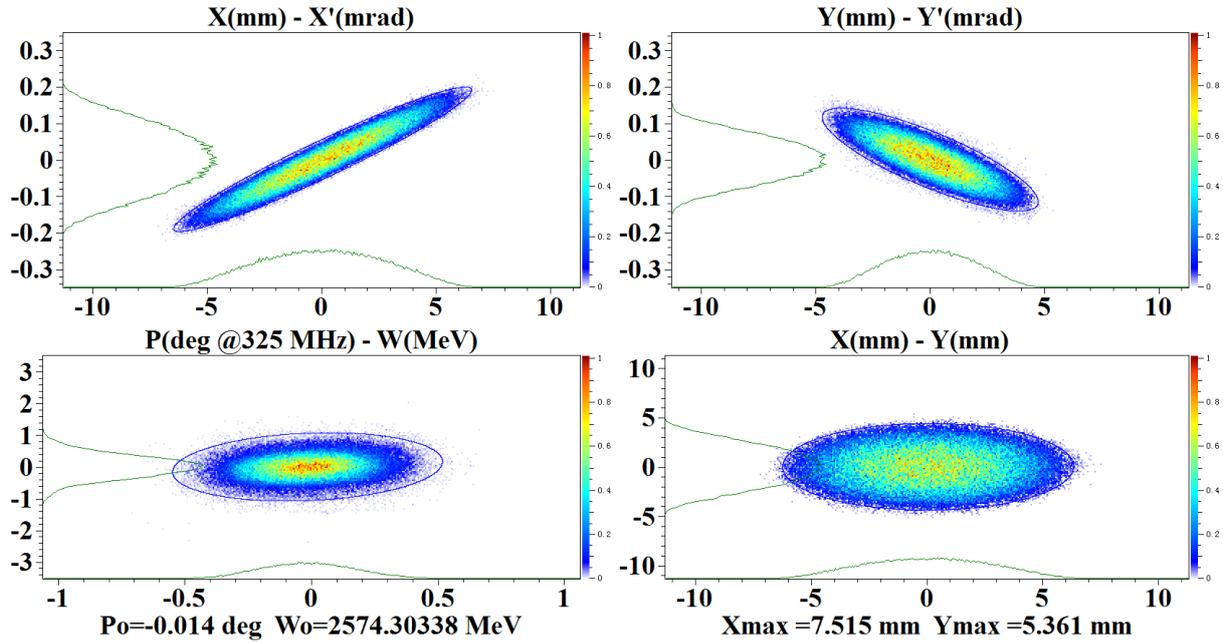

Fig 8 phase space distributions at the end of the linac (2.5 GeV)

We have also investigated the possibility of introducing a new type of high-beta (0.93) elliptical cavity to cover the high-energy section 1.0-2.5 GeV to save the total cavity number and the cost. The preliminary results show that only 15 cavities and about 20 meters are saved. As this scheme asks for developing a new cavity type which demands additional R&D efforts, the advantage is considered marginal.

Of course, the design presented here is still preliminary. Along with the progress of the C-ADS project, there will be more and more inputs from the hardware developments and the design will be further optimized.

## 4  Conclusions

A 15-MW CW superconducting proton linac is designed as the driver for the proposed China neutrino beam facility - MOMENT. The preliminary beam dynamics study shows that the proposed design scheme can satisfy the requirements of the project. Of course, for such a very high power proton linac, there are many technique difficulties which need to be solved through intense R&D efforts, for example, the RFQ working in CW operation mode, low-beta superconducting cavities, high-power and large-scale RF amplifiers, cryomodules with many elements and high average heat load, very strict beam loss control and so on. Fortunately, the China-ADS project is executing a strong R&D plan and the experimental facility to be built in about 10 years is expected to solve the major problems. As the progress of the C-ADS project, more and more engineering inputs obtained from the R&D will be integrated into the physics design and the scheme will be further optimized.

## References

[1] Y.F. Wang, Neutrino physics and requirements to accelerators, Proc. of IPAC2013, Shanghai, China, (2013) p.4010
[2] Jingyu Tang, A medium baseline neutrino superbeam in China, 2013 International Workshop on

   Neutrino Factory, August 19-24, Beijing
[3] Zhihui Li et al., Phys. Rev. ST Accel. Beams 16, 080101 (2013)
[4] R.L. Sheffield, in Proceedings of HB2010, Morschach, Switzerland, 2010(PSI, Viligen, Switzerland, 2011), pp. 1-5.
[5] LI Zhi-hui, TANG Jing-Yu, YAN Fang, et al., Study of the longitudinal instability caused by long drifts in the C-ADS injector-I, Chinese Physics C, Vol. 37, No. 3 (2013)037005
[6] F. Gerigk, Proceedings of the 2002 Joint USPAS-CAS-Japan-Russia Accelerator School (2002), pp. 257–288.
[7] I. Hofmann, Phys. Rev. E 57, 4713 (1998).5
[8] K.R. Crandall et al., 1998, RFQ Design Codes, LANL Report LA-UR-96, p. 1836
[9] http://irfu.cea.fr/Sacm/logiciels/index3.php.